\documentclass[sigconf]{acmart}
\renewcommand\footnotetextcopyrightpermission[1]{} 
\usepackage{float}
\usepackage{amsmath,amssymb}
\usepackage{bm}
\usepackage{microtype}

\newcommand{\bbeta}{\bm{\beta}}
\newcommand{\bX}{\bm{X}}
\newcommand{\by}{\bm{y}}
\newcommand{\btheta}{\bm{\theta}}
\newcommand{\bA}{\bm{A}}
\newcommand{\bI}{\bm{I}}
\newcommand{\ba}{\bm{a}}
\newcommand{\bPhi}{\bm{\Phi}}

\def\BibTeX{{\rm B\kern-.05em{\sc i\kern-.025em b}\kern-.08emT\kern-.1667em\lower.7ex\hbox{E}\kern-.125emX}}
    
\copyrightyear{2018}
\acmYear{2018}
\setcopyright{acmlicensed}

\acmConference[]{}{}{}
\acmYear{2019}
\copyrightyear{2019}

\acmArticle{4}
\acmPrice{15.00}

\begin{document}
%
\title{Three Methods for Training on Bandit Feedback}

\author{Dmytro Mykhaylov}
\affiliation{%
   \institution{Criteo Research}
   \city{Paris}
}
\email{d.mykhaylov@criteo.com}

\author{David Rohde}
\affiliation{%
   \institution{Criteo Research}
   \city{Paris}
}
\email{d.rohde@criteo.com}

\author{Flavian Vasile}
\affiliation{%
   \institution{Criteo Research}
   \city{Paris}
}
\email{f.vasile@criteo.com}

\author{Martin Bompaire}
\affiliation{%
   \institution{Criteo Research}
   \city{Paris}
}
\email{m.bompaire@criteo.com}

\author{Olivier Jeunen}
\affiliation{%
   \institution{University of Antwerp}
   \city{Antwerp}
}
\email{olivier.jeunen@uantwerpen.be}

%
%
\begin{abstract}
	There are three quite distinct ways to train a machine learning model on recommender system logs.  The \textit{first} method is to model the reward prediction for each possible recommendation to the user, at the scoring time the best recommendation is found by computing an argmax over the personalized recommendations.  This method obeys principles such as the conditionality principle and the likelihood principle. A \textit{second} method is useful when the model does not fit reality and underfits. In this case, we can use the fact that
	we know the distribution of historical recommendations (concentrated on previously identified good actions with some exploration)  to adjust the errors in the fit to be evenly distributed over all actions.  \textit{Finally}, the inverse propensity score can be used to produce an estimate of the decision rules expected performance.  The latter two methods violate the conditionality and likelihood principle but are shown to have good performance in certain settings.  In this paper we review the literature around this fundamental, yet often overlooked choice and do some experiments using the \textit{RecoGym} simulation environment.	
\end{abstract}
%
%

	\ccsdesc[500]{Computing methodologies~Supervised learning by regression}
\keywords{Bandit algorithms, Contextual Bandits, Bayesian Methods}


%
\keywords{}

%

%
\maketitle

\section{Three methods for estimating the short term reward of actions }

Imagine that we have logs of a recommender system that takes users with context $\bX$, delivers recommendation or action $\ba$, and receives a short term reward as a click or no click $c$.

We may approach this problem from three distinct directions.  One method respects the likelihood principle, the conditionality principle and can be consistent with the Bayesian axioms is to build a likelihood-based model of the form:

\[
c_n \sim {\rm Bernoulli}
  \left(
    \sigma
      \left(
	\bPhi {\left( [\bX_n ~ \ba_n] \right)}^T \bbeta
      \right)
  \right)
\]

\noindent
where
\begin{itemize}
  \item $\bbeta$ are the parameters;
  \item $\sigma \left( \cdot \right)$ is the logistic sigmoid;
  \item $\bPhi \left( \cdot \right)$ is a function that maps $\bX,\ba$ to a higher dimensional space and includes some interaction terms between ${\bX}_n$ and ${\ba}_n$ - without interaction terms there is no personalization and recommendation would fall back to best-of.  We assume that the action is discrete and ${\ba}$ uses one-of-n coding.
\end{itemize}

A simple cross-product is sufficient to get some level of personalization i.e. $\bPhi \left( [\bX_n ~ \ba_n] \right) = \bX_n \otimes \ba_n$, where $\otimes$ is the Kronecker product.

The estimation of this model can be achieved using Bayesian methods, or more conveniently using the principal of maximum likelihood:

\begin{align*}
	\hat{\bbeta}_{\rm lh} = {\rm argmax}_{\bbeta} &
	\sum_n c_n \log \sigma
		\left(
			{
				\bPhi
					\left(
						\left[ {\bX}_n ~ {\ba}_n \right]
					\right)
			}^T \bbeta
		\right) \\
	& + \left( 1-c_n \right)
	\log
	\left(
		1 - \sigma
			\left(
				\bPhi
					{\left(
						\left[
							{\bX}_n ~ {\ba}_n
						\right]
					\right)}^T \bbeta
			\right)
	\right)
\end{align*}.

Alternatively using $\bPhi \left( \left[ {\bX}_n ~ {\ba}_n \right] \right) = {\bX}_n \otimes {\ba}_n$:

\begin{align*}
	\hat{\bbeta}_{\rm lh} = {\rm argmax}_{\bbeta}
	& \sum_n c_n \log \sigma
				\left(
					{
						\left(
							{\bX}_n \otimes {\ba}_n
						\right)
					}^T \bbeta
				\right) \\
	& + \left(
		1-c_n
	  \right)
	  \log
	  \left(
	  	1 - \sigma
			\left(
				{
					\left(
						{\bX}_n \otimes {\ba}_n
					\right)
				}^T \bbeta
			\right)
	\right)
  \end{align*}

This model allows (greedy) recommendations to be delivered by computing:

\begin{align}
	{\ba}_n^* = {\rm argmax}_{{\ba}_n}
		{\bPhi
			\left (
				\left[
					{\bX}_n ~ {\ba}_n
				\right]
			\right)}^T \bbeta_{\rm lh} 
\label{score}
\end{align}

This model is a direct application of logistic regression using standard maximum likelihood. Naturally, it respects the conditionality and likelihood principles which we define in detail in Section~\ref{likelihood}.

Alternate approaches are usually motivated from the observation that the logs are typically unbalanced, i.e. the recommender system will typically have set the historical value of $\ba$ to maximize historical clicks.  If the model lacks capacity, a notion that we will make more precise in Section~\ref{rew} then this may introduce a bias where the estimate for rarely used (typically poor) recommendations will be sacrificed in order to estimate the click-through rate of more common actions, however when we determine the best action  the model is evaluated for each action (i.e. uniformly on actions), this is a version of the so-called \emph{domain shift} problem \cite{shimodaira2000improving} \cite{sugiyama2006importance}.  A significant literature has developed around domain shift, and the most basic proposal is to use a re-weighting that adjusts for the difference in the distribution of past actions (as per the policy $\pi(a|x)$) and future actions (which we will evaluate uniformly in this case).  This results in a weighted logistic regression problem:

\begin{align*}
	\hat{\bbeta}_{\operatorname{re-weight}} =  {\rm argmax}_{\bbeta}
	\sum_n
	& w_n c_n \log \sigma
		\left(
			{\bPhi
				\left(
					\left[ {\bX}_n ~ {\ba}_n \right]
				\right)
			}^T \bbeta
		\right)  \\
	& + w_n \left( 1-c_n \right)
		\log
			\left(
				1 - \sigma
					\left(
						{\bPhi \left(
								\left[ {\bX}_n ~ {\ba}_n \right]
							\right)
						}^T \bbeta
					\right)
			\right)
\end{align*}

or

\begin{align*}
	\hat{\bbeta}_{\operatorname{re-weight}} =
	{\rm argmax}_{\bbeta}
	\sum_n
	&       w_n c_n 
		\log 
		\sigma
			\left(
				{\left(
					{\bX}_n \otimes {\ba}_n
				\right)}^T
				\bbeta
			\right) \\
	& 
		+ w_n \left( 1 - c_n \right)
		\log
		\left( 1- \sigma \left(
					{\left(
						{\bX}_n \otimes {\ba}_n
					\right)}^T \bbeta
				  \right)
		\right)
\end{align*}

\noindent
where the weight is defined: $w_n = \frac{1}{\pi\left( {\ba}_n | {\bX}_n \right)}$.

Another variant often called a \textit{contextual bandit} is based on the notion of off policy training.  Previously we produced a personalized model of the probability of a click for every possible action and then maximized this to find the best action.  Instead, we might merely propose a new decision rule or policy which is a mapping from the user context $\bX$ directly to an action $\ba$. This class of methods pose a new probabilistic policy $\pi_{\bbeta}\left( {\ba}_n | {\bX}_n \right)$ even though the optimal policy will be degenerate this probabilistic formulation allows the use of importance sampling to evaluate the ``counterfactual risk''.

\[
	\pi_{\bbeta}\left( {\ba}_n | {\bX}_n \right) = {\rm softmax}
		\left( {\bPhi
			\left(
				\left[ \bX_n ~ \ba_n \right]
			\right)}^T \bbeta
		\right)
\]

\noindent
or

\[
	\pi_{\bbeta} \left( {\ba}_n | {\bX}_n \right) = {\rm softmax}
		\left(
			{\left(
				{\bX}_n \otimes {\ba}_n
			\right)}^T \bbeta
		\right)
\]

\[
	\hat{V} = \frac{1}{N}\sum_n^N \frac{
		c_n \pi_{\bbeta}
		\left(
			{\ba}_n | {\bX}_n
		\right)
	}{
		\pi
		\left(
			{\ba}_n | {\bX}_n
		\right)
	} = \frac{1}{N} \sum_n^N w_n c_n \pi_{\bbeta}
	\left(
		{\ba}_n | {\bX}_n
	\right)
\]

\noindent
Here $\hat{V}$ is an estimator of the expected number of clicks if the policy $\pi_{\bbeta}\left(\cdot\right)$ is used.  Optimizing this quantity directly requires custom software.  Alternatively the problem can be modified to make it resemble weighted maximum likelihood, this can achieved if we bound the counterfactual risk using Jensen's inequality:

\begin{align*}
\log \left( \sum_n^N w_n c_n \pi_{\bbeta} \left({\ba}_n | {\bX}_n \right) \right)  & \ge  \frac{\sum_n^N  w_n c_n \log \pi_{\bbeta} \left({\ba}_n| {\bX}_n\right) } {\sum_n w_n c_n} + \log(\sum_n^N w_n c_n)
\end{align*}

\noindent
Which can be maximized simply by maximizing the following familiar form of weighted multi-class logistic regression: 

\[
\sum_n^N w_n c_n \log \pi_{\bbeta} \left({\ba}_n| {\bX}_n\right)
\]

\noindent
Optimizing the lower bound then has the same form as a weighted multiclass log likelihood giving the contextual bandit optimization problem:

\begin{align*}
	\hat{\bbeta}_{\rm CB} = {\rm argmax}_{\bbeta} \sum_n
	& w_n c_n
		{\left(
			{\bX}_n \otimes {\ba}_n
		\right)}^T ~ \bbeta  - w_n c_n \log \sum_{{\ba}_{n}^{'}} e^{
						{\left(
							{\bX}_n \otimes {\ba}_{n}^{'}
						\right)}^T \bbeta
						} \\
\end{align*}

\noindent

Another argument for using weighted multiclass classification (without clearly specifying if it is the probability or the log probability to be weighted) is given in \cite{beygelzimer2009offset} under the name ``one-step reinforcement learning reduction''.

It is an important remark that a direct optimization of counterfactual risk rather than the bound does not resemble (weighted) log likelihood because it involves a weighted sum of distributions rather than a weighted sum of log distributions.

\section{Discussion of the three methods}

We present such a parameterization that all three methods use the same dimensional parameter of $\bbeta$, and in all three cases actions are found using $\ba_n^* = {\rm argmax}_{\ba_n} {\left( {\bX}_n \otimes {\ba}_n \right)}^T \bbeta$.  This common parameterization is useful for allowing the three methods to be compared.

\subsection{Likelihood} \label{likelihood}
The use of likelihood is based on the fact that theoretically under standard modeling assumptions there is no need to make adjustments to the estimation based on covariate shift.  This is simply a consequence of assuming the following factorization where there is no covariate shift on ${\bX}_{N+1}$:

\begin{align*}
  \Pr
	& \left(
		{\by}_{1:N},{\bX}_{1:N},y_{N+1},{\bX}_{N+1}
	\right) = \\
  	& \int \int \prod_{i=1}^{N+1}
		\Pr	\left(
				y_i|{\bX}_i,\bbeta
			\right)
		\Pr	\left(
				{\bX}_i|\btheta
			\right)
		\Pr	\left(
				\bbeta
			\right)
		\Pr	\left(
				\btheta
			\right)
		d\bbeta d\btheta,
\end{align*}

\noindent
and this distribution where there is a covariate shift on ${\bX}_{N+1}$

\begin{align*}
  \Pr
	&	\left(
			\by_{1:N},\bX_{1:N},y_{N+1},\bX_{N+1}
		\right) =  \\
  	& \int \int \int
		\left(
		\prod_{i=1}^{N}
			\Pr
				\left(
					y_i|{\bX}_i,\bbeta
				\right)
			\Pr
				\left(
					{\bX}_i|\btheta
				\right)
		\right) \\
  & \times
	\Pr
		\left(
			y_{N+1}|{\bX}_{N+1},\bbeta
		\right)
	\Pr
		\left(
			{\bX}_{N+1}|\btheta^*
		\right)
	\Pr \left(\bbeta\right)
	\Pr \left(\btheta\right)
	\Pr \left(\btheta^*\right)
	d\bbeta
	d\btheta
	d\btheta^*;
\end{align*}

\noindent
both have the same conditional distribution i.e.

\begin{align*}
  \Pr
	&	\left(
			y_{N+1}|{\bX}_{N+1},\by_{1:N},\bX_{1:N}
		\right) =  \\
  	& \int \Pr	\left(
				y_{N+1}|\bX_{N+1},\bbeta
			\right)
		\Pr	\left(
				\bbeta|{\by}_{1:N},{\bX}_{1:N}
			\right)
		d\bbeta.
\end{align*}

\noindent
This argument originates from \cite{storkey2009training}.

A further argument in favor of likelihood is that it does not use the inverse propensity score.  A method that uses the inverse propensity score violates the conditionality principle and as a consequence also the likelihood principle \cite{birnbaum1962foundations} \cite{berger1988likelihood}.  The conditionality principle is the principle that experiments that were not performed are irrelevant, the likelihood principle is implied by the combination of the conditionality principle and the sufficiency principle.  The likelihood principle states that the likelihood function contains all relevant information for inferences, a direct implications of these principles is the irrelevance of the IPS.

Methods that use IPS based estimation are part of a long tradition of methods that adopt estimators that have good properties, the error rates of the estimator are established before the analysis takes place.  In contrast, the conditionality principle demands that error rates are computed after the data becomes available. 

One of the most celebrated example of the likelihood principle (e.g. see Chapter 2, Example 2 in \cite{berger1988likelihood}) involves a coin flip being used to determine if an accurate or inaccurate measuring device should be used.  If the entire system is analyzed, then the error rate should incorporate the coin flip and average over both possibilities.  That is we should model the error from the measuring device as a mixture of the two experiments that may have occurred instead of the one that did occur.  This is against most peoples intuition and the conditionality principle which states that error rates should be reported for the measuring device that is actually used.

An analogous situation occurs in a recommender system that uses the IPS.  Instead of considering the actions that actually were performed by the recommender system, the estimator uses long run arguments averaging over the typical behavior of the system.

We do not necessarily think such theoretical or philosophical arguments should be decisive for the recommender systems community, but it is useful to be aware of them.

A potential downside of applying likelihood or Bayesian methods is that the estimation is done before the decision is made.  This means that these methods try to estimate the outcome for every possible action equally well; we will see shortly this is not the case for other methods.  Indeed the estimation method does not know that you are trying to find the action that has the highest reward and would be identical if you wanted to find the action with the lowest reward.

Also note some explore strategies such as Thompson sampling and Upper Confidence Bound require uncertainties that naturally arise from a Bayesian framework consistent with the likelihood principle and the conditionality principle.

\subsection{Re-weighted Likelihood}
\label{rew}

A significant literature has developed around the term covariate or domain shift for re-adjusting likelihood-based estimation based on the fact that the distribution has shifted during training.  Numerous studies have shown the apparent benefits of these methods  e.g. \cite{shimodaira2000improving} \cite{sugiyama2006importance}.

\begin{figure}[h]
  \centering
  \includegraphics[width=0.8\linewidth]{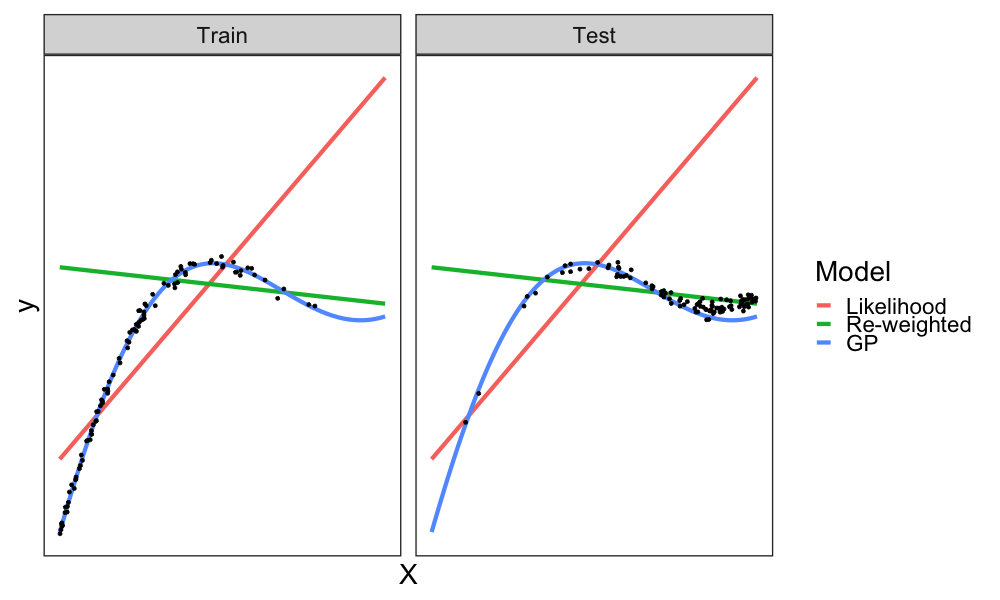}
  \caption{The effect of estimating a linear model of a non-linear effect both with and without IPS re-weighting.  The IPS re-weighting function causes the underfit section to be reduced on the target domain.  A flexible non-linear model the Gaussian process (Gaussian process) can perform well at all points on the function.}
  \label{domainshift}
\end{figure}

The apparent success of these methods is explained in \cite{storkey2009training} as being due to underfitting.  
This is due to a model not having sufficient capacity to map the model's response to each input to any arbitrary value.  If a model lacks capacity, then the maximum likelihood will be biased to regions where there is more data at the expense of regions where data is sparse.  An IPS based adjustment can, therefore, be used to re-balance the errors so they are minimized in the target domain, in the case of a recommender system the target domain would be for uniform actions.  This phenomenon is illustrated in Figure~\ref{domainshift} which demonstrates a linear model fit to a non-linear function using both an IPS based re-weighting and pure likelihood, as expected the IPS based re-weighting improves performance in the target domain, also as expected a non-linear model is able to perform well everywhere.

The likelihoodists argument against re-weighting is simply to use a model that does not underfit.  The proponent of re-weighting would respond by saying restricting the parameter space is important for reducing the variance of statistical estimation on finite samples, this is certainly true for methods applying maximum likelihood but untrue of the other broad class of likelihood based methods: Bayesian methods.

Both likelihood and re-weighted likelihood involve evaluating the model over every possible action.  Taking the maximum of many noisy estimates can exacerbate the impact of noise a phenomenon known as the optimizer's curse \cite{smith2006optimizer}.  An erroneously high estimate of the reward for any single action can be very detrimental to performance.

\subsection{Contextual Bandits}

A contextual bandit \cite{li2010contextual} uses quite different principles in that they do not optimize a model, rather they optimize a decision rule or policy that directly maps the user context to an action\footnote{In this paper we are interested in training on ``logged bandit feedback'' we are not interested in the explore exploit dilemma which  the word bandit sometimes evokes.}.  This can be thought of as using the model's capacity exclusively to determine good actions, and capacity is not wasted in ordering the worst and second worst action.  The mixing of the model and the decision rule into a single optimization problem can introduce increased risk of overfitting as apparently highly optimized solutions may arise from assuming that nature is more favorable than it is.

There are two arguments in the literature that the IPS must be used in order to obtain good long run behavior.  One of these argument originates in \cite{robins1997toward} and has been colorfully covered in a sequence of blog posts by Larry Wasserman and James Robbins titled ``Robins and Wasserman Respond to a Nobel Prize Winner'' with several responses by the Nobel laureate in question Christopher Sims.  The argument is posed in statistical terms, that do not necessarily make its connection to recommender systems clear. Indeed the use of personalization or large recommendation sets is not required to illustrate their example.  The argument is relevant to estimating the reward of a single action without personalization, i.e. we are interested in the reward ($\alpha$) for giving action or recommendation $\ba^*$ to the entire population:

\[
	\alpha = \mathbb{E}_{\bX} \left[\Pr \left( c|\bX,\ba^* \right)\right]  
\]

Theorem 3  in \cite{robins1997toward} shows that it is not possible to estimate $\alpha$ uniformly consistently unless the estimator uses the propensity score, i.e. $\Pr(\bA^*=\ba^*)$ here, hence violating the likelihood principle.  The responses of Sims, who advocates for likelihood based methods also uses the propensity score, so in violating conditionality they are apparently in agreement (On this basis Robbins and Wasserman, quite reasonably, question if Sim's method is really Bayesian).  Further discussion of this example is given in: \cite{robinsrobins} \cite{sims2006example} \cite{robins2000conditioning} \cite{asimple2012robins}.

A related example more directly related to the recommendation is given in \cite{beygelzimer2009offset}, where it is argued that a likelihood-based training algorithm (or any value based method) have a regret that is bounded by the square root of the regressor's regret.  Regret is defined as the difference in loss between the action taken and the best possible action.  Their analysis shows better regret performance for contextual bandit based algorithms.  

Another line of argument concerns not statistical properties but simply asks the question:
\emph{is including a variable sufficient to infer a correct causal relationships?}.  A balancing score is a function of the covariate such that the probability of their treatment is the same for each action, while the full covariate $\bX$ is a balancing score so is the propensity score  \cite{rosenblum1983central}, as a consequence IPS can be used to get a consistent estimate.

\subsection{Another Likelihood-based approach: Bayesian methods}

In the previous sections, we demonstrated that using the IPS score had advantages in terms of capacity.  The re-weighted method uses the best use of insufficient capacity by sharing the error evenly over the outcome of all possible recommendations, and the contextual bandit is even more extreme in the sense that it merely attempts to find the best recommendation for each context using its capacity directly on the decision rule.

The likelihood-based method seems to rely on the model being sufficiently flexible to model reality; however statistical point estimation over complex models can require many more samples than for simpler models.  This makes it difficult for maximum likelihood to compete with the IPS based methods at small samples.  There are however two broad families of likelihood-based methods, maximum likelihood and Bayesian methods.  When sophisticated priors are used Bayesian methods can perform well even when the model is complex, and the sample is small: this is because Bayesian methods do not overfit. Indeed they do not ``fit'' but rather condition on the data.  We suggest that a reasonable prior over $\bbeta$ is:

\[
	\bbeta \sim \mathcal{N}
		\left(
			\mu,
			\left(a\bI + b\right)
			\otimes
			\left(a\bI + b\right)
		\right)
\]

\noindent
The covariance matrix $(a\bI + b) \otimes (a\bI + b)$ has a checkerboard pattern consisting of three values.  $(a+b)^2$ on the diagonal, $(a+b)b$ if there is either a history item or an action in common and $b^2$ otherwise.  In terms of correlation, it is $b/(a+b)$ between click-through rates with either an action or a history in common, or  $b^2/(a+b)^2$ otherwise.  It is natural to suppose that similar items i.e. those that share either an action or a history item are more similar, it may be less clear why we want to correlate unrelated actions.  The reason for this is that even unrelated actions help establish the range of plausible click-through rates.

This model is similar to the model given in \cite{smith2006optimizer} but rather than using a uniform prior on the response, the prior pulls the click through rates towards each other especially those that are similar. 

\section{Experiments}

We use the \textit{RecoGym} simulation environment \cite{rohde2018recogym} to simulate A/B tests with different numbers of samples for each of the three classes of methods.  Our experimental setup involves the usual
recommendation setup of training on an offline logs and then using this model to produce a model that we will deploy in production.  We consider two types of offline logs:


\paragraph{Popularity-based}
A simple yet effective baseline policy is to sample actions with probabilities proportionate to the occurrence frequency of the item in the user's historical interactions.  Because this is a reasonable baseline policy we consider it as a logging policy.  

The popularity based policy operates in the following way.  Imagine a simplified setting with 3 products and a user state of $[3,1,0]$, this means we sample item $1$ with probability $\frac{3}{4}$, item $2$ with probability $\frac{1}{4}$, and we don't sample item $3$.
In general, for a user history $\mathbf{x}$, $\pi_\text{pop}(a|\mathbf{x}) = \frac{\mathbf{x}_a}{\sum_{i=1}^{n}\mathbf{x}_i}$.
This policy does not have full support over the item catalogue, violating the assumptions that guarantee importance sampling to yield an unbiased estimate~\cite{Owen2013}, it does however reflect standard behavior of a recommender system.

\paragraph{Inverse popularity-based}
Although it may not be the most natural approach it is academically interesting also to consider a logging policy that historically mostly selected poor actions.
In order to achieve this we invert the probabilities obtained through the popularity-based policy and renormalize.
That is, $\pi_\text{inv-pop}(a|\mathbf{x}) = \frac{1 -\pi_\text{pop}(a|\mathbf{x})}{\sum_{i=1}^{n}1 - \pi_\text{pop}(a|\mathbf{x})}$.


Within the likelihood based methods we test maximum likelihood as well as Bayesian methods with various hyper-parameter settings.
This results in testing eight \textit{RecoGym} agents in total.  We run the simulator for $2000, 4000, 6000$ and $8000$ samples in order to evaluate the behavior of the method with both large and small samples.  After training an A/B test is simulated on $10000$ samples.  There is no explore exploit aspect being studied here we simply investigate how each method performs with varying sample sizes.
All \textit{RecoGym} settings are set to their defaults which results in a recommendation problem with a catalogue of size 10 clustering into two main classes of product.  For all experiments, we set $\mu=-6$ noting that $\sigma(-6) \approx 0.0025$ reflecting a prior belief that click-through rates are often low especially for bad recommendations, we tried several values of hyperparameters $a$ and $b$ and found that $a=b=0.01$, the Bayesian solution was very sensitive to these parameters and for a poor choice it was systematically beaten by all other methods.

\begin{figure}[h]
  \centering
  \includegraphics[width=0.6\linewidth]{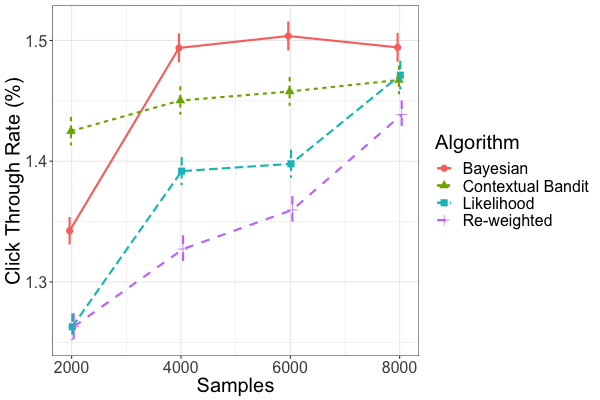}
  \includegraphics[width=0.6\linewidth]{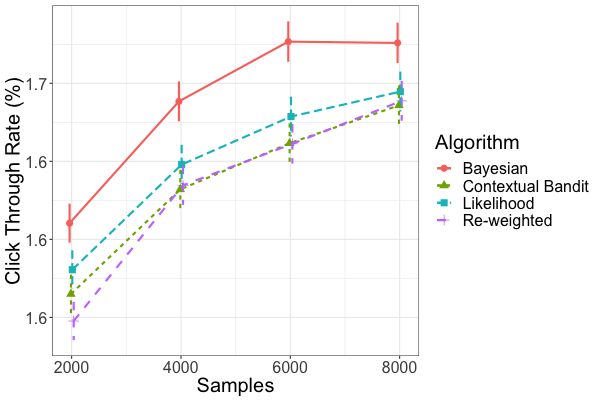}
	\caption{The performance of the three methods as a function of the number of samples assuming the logging policy is popularity based (above) or inverse popularity based (below).}
\label{results}
\end{figure}


In Figure~\ref{results} (top) the actions of the logged data are drawn according to item popularity.  Out of the three-point estimation methods, likelihood, re-weighted likelihood and the contextual bandit we see that the contextual bandit performs the best at low samples and remains competitive for high samples.  The Bayesian methods have two additional hyper-parameters $a$ and $b$.  We observed the performance is quite sensitive to these values, we obtained the best performance for $a=b=0.01$ where the Bayesian method has the highest click-through rate beating the contextual bandit by 0.005, although the performance is less than the  contextual bandit for 2000 samples.

In Figure~\ref{results} (bottom) the actions of the logged data are drawn according to inverse item popularity.  Here the Bayesian method consistently wins, likelihood also outperforms the contextual bandit.  The re-weighted model consistently performs poorly in both cases.



\section{Conclusion}

We reviewed different ways of training machine learning models on recommender system logs.  Complex theoretical and philosophical arguments sit behind choices that recommender system practitioners must make.  We used the \textit{RecoGym} environment to investigate how the three approaches performed with different sample sizes, when the data was biased towards good recommendation we saw better low sample performance from the contextual bandits, this advantage vanished in the (admittedly) artificial case of inverse popularity sampling.  Bayesian methods with appropriate priors performed well in all settings (except the 2000 sample popularity sampling case).  

\bibliographystyle{plain}
\bibliography{literature}

\end{document}